*Article*

# Quarantining Malicious IoT Devices in Intelligent Sliced Mobile Networks

**David Candal-Ventureira** [†], **Pablo Fondo-Ferreiro** [†], **Felipe Gil-Castiñeira** [*]
and **Francisco Javier González-Castaño**

Information Technologies Group, atlanTTic Research Center for Telecommunication Technologies,
University of Vigo, 36310 Vigo, Spain; dcandal@gti.uvigo.es (D.C.-V.); pfondo@gti.uvigo.es (P.F.-F.);
javier@gti.uvigo.es (F.J.G.-C.)
* **\*** Correspondence: xil@gti.uvigo.es; Tel.: +34-986-818-665
* † These authors contributed equally to this work.

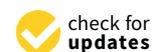

**Abstract:** The unstoppable adoption of the Internet of Things (IoT) is driven by the deployment of new services that require continuous capture of information from huge populations of sensors, or actuating over a myriad of "smart" objects. Accordingly, next generation networks are being designed to support such massive numbers of devices and connections. For example, the 3rd Generation Partnership Project (3GPP) is designing the different 5G releases specifically with IoT in mind. Nevertheless, from a security perspective this scenario is a potential nightmare: the attack surface becomes wider and many IoT nodes do not have enough resources to support advanced security protocols. In fact, security is rarely a priority in their design. Thus, including network-level mechanisms for preventing attacks from malware-infected IoT devices is mandatory to avert further damage. In this paper, we propose a novel Software-Defined Networking (SDN)-based architecture to identify suspicious nodes in 4G or 5G networks and redirect their traffic to a secondary network slice where traffic is analyzed in depth before allowing it reaching its destination. The architecture can be easily integrated in any existing deployment due to its interoperability. By following this approach, we can detect potential threats at an early stage and limit the damage by Distributed Denial of Service (DDoS) attacks originated in IoT devices.

## 1. Introduction

The 5G paradigm aims at defining "universal networks" connecting devices of all kinds. Billions of devices with different needs in terms of throughput, latency, reliability, security and density of users will be connected through 5G networks. The Internet of Things (IoT) is explicitly considered by the 3rd Generation Partnership Project (3GPP) in the design of new networks [1]. In fact, massive IoT is one of the use cases that has attracted more interest from standardizing bodies. Sensors and actuators servicing critical infrastructures such as energy, health, vehicles and telecommunications are expected to be connected through 5G networks. These devices exchange highly sensitive information, thus wrong manipulation of their communications may have significant economic, security and social impact.

IoT devices have very different characteristics and requirements compared to traditional smartphones or other devices under the enhanced Mobile Broadband (eMBB) class, which form the majority of legacy mobile network terminals. They are simple devices with low computing power, so they are highly vulnerable to security issues. They usually have low encryption capabilities and only support simple authentication procedures [2]. Due to their low processing power, they may require computational offloading to a cloud computing server, further favoring security breaches. Moreover, even though IoT devices are prone to issues that may not only put the device itself but also the whole network connecting it at risk, organizations are not showing too much interest in patching them [3].

Since their origins, cellular networks have been the target of numerous attacks. These increased in number with the deployment of the fourth generation of mobile technology. The IP-based architecture of 4G core networks led to the emergence of attack vectors which had never been seen before, whereas the increasing dependence on cellular networks also raised the interest of the attackers. The introduction of new use cases in 5G networks will foster this interest. In fact, IoT Distributed Denial of Service (DDoS) attacks are envisioned as one of the worst security threats 5G networks will need to face. IoT botnets have reached a high degree of maturity, up to the point that the most powerful and frequent DDoS attacks are performed by IoT botnets conformed by vulnerable commodity devices [3,4]. Therefore, operators must implement mechanisms to detect these attacks and perform countermeasures to protect the rest of the network against them.

5G networks are being designed around new concepts and paradigms that may be exploited to improve their security. Software-Defined Networking (SDN) [5–7] allow reconfiguration based on complete network information, as well as assisting in the creation of consistent policies. Network Function Virtualization (NFV) [8–10] enables the deployment of network elements over a generic computing infrastructure. NFV enhances the availability of mobile network elements, as the operator is empowered to configure the network and the computational capabilities of each node based on its necessities, as well as deploying backup nodes in case the main ones become compromised. Network Slicing (NS) [11–15] supports the creation of multiple complete end-to-end networks providing different use cases independently. By isolating the traffic flows of these different use cases it may be easier to ensure the Quality of Service (QoS) requirements in the Service Level Agreements (SLAs) of end users, as well as to provide an additional security layer among the different connected devices.

In this paper we propose a novel architecture for an efficient detection of DDoS attacks originating from massive IoT devices in mobile networks. Instead of applying Deep Packet Inspection (DPI) to all the data flows in the network, our solution, which relies on the SDN, NFV and Network Slicing concepts, applies a two-step detection procedure. A first detection procedure, executed by a lightweight SDN application, performs a coarse-grained identification of suspicious sets of end devices which are believed to include malicious nodes. The data flows of the suspicious nodes are diverted to a quarantine slice, in which they can be analyzed in depth to set malicious devices apart from legitimate ones. This two-step process reduces the computational load of the DDoS detection system of the mobile network, since traffic is only inspected in depth on demand on a reduced set of data flows.

Summing up, the main contributions of this paper are:

- A two-step SDN-based architecture for detecting and mitigating DDoS attacks in IoT cellular networks.
- The analysis of the detection capabilities of the proposed solution through simulations.
- The evaluation of the reassignment of IoT devices to quarantine slices in a real-world testbed.

The rest of the paper is organized as follows. Section 2 discusses related work. Our proposed solution is described in Section 3 and evaluated in Section 4. Finally, Section 5 concludes the paper.

## 2. Related Work

DDoS attacks from IoT botnets are a main concern in 5G networks security. SDN, one of the most significant paradigms in 5G core network design, has different advantages to build better defense systems for modern networks [16]. Its global vision of the network and its flexibility in routing reconfiguration simplifies the development of complex security systems aimed at efficiently detecting and mitigating attacks and failures in the network [17].

There is abundant work in the literature on the detection of DDoS attacks on networks by relying on SDN. In [18], the authors propose a framework for improving network security in which data traffic is mirrored to a central Intrusion Detection System (IDS) for attack detection, taking into account the mobility of the users. In [19,20], two machine learning models for detecting malicious data flows are presented. In [21], a joint entropy-based statistical model classifies ongoing data flows as legitimate or malicious ones in case of congestion. The importance of detecting DDoS attacks as close as possible to the attackers themselves, both to detect them more quickly and to reduce wasted backbone network resources, is discussed in [22]. The analysis of the data flows is taken to the switches themselves in [23], by relying on the concept of programmable data plane. This way, control plane traffic is reduced at the expense of requiring more complex network devices. This work also introduces a mechanism to prevent DDoS attacks against SDN switches by reducing the size of the flow tables of the switches using wildcards.

These works take advantage of the global knowledge of the SDN controller for detecting malicious devices; the centralized programmability of the network control plane, which permits creating consistent policies to drop all the data frames that match certain filters; or the sampling capabilities of SDN networks, which allow mirroring part of the traffic towards an analysis entity. Although the delay in acting on malicious traffic can increase the severity of the problem, immediate blocking of all suspicious traffic can lead to the disruption of legitimate user services since false positives may occur. In [24] the concept of quarantine slice was introduced for the first time. The idea behind this term is to isolate suspicious traffic by moving the corresponding flows to a new network slice with restricted capacities. This way, suspicious traffic can be further analyzed before blocking its sessions without affecting legitimate traffic. In [25] a quarantine slice is used to isolate suspicious flows. However, this work focuses on the detection of malicious nodes (mainly by analyzing network slice requests) at a high level and it does not propose any architecture for diverting malicious flows to a quarantine slice.

Suspicious traffic isolation is relevant not only at academic level but also at industrial level. Research projects [26] and patents [27] include the concept of quarantine slice for the aforementioned purposes. The ANASTACIA project (http://www.anastacia-h2020.eu/) seeks to develop a solution for a network serving IoT devices to act automatically and intelligently in the face of threats. This project introduces an architecture to move users to a quarantine slice if suspicious behavior is detected. However, anomaly detection is carried out by specific entities that perform deep packet inspection of user traffic to determine if a flow is potentially malicious.

In this paper we present a novel two-step DDoS attack detection solution for current mobile networks supporting massive IoT use cases. A coarse-grained detection procedure running in an SDN application is in charge of detecting abnormal growth of throughput from aggregated sets of IoT devices. Aggregation of IoT devices allows coping with the flow table limits of SDN switches, so that it becomes possible to handle the vast number of devices that 5G networks will support in massive IoT scenarios. In case a set of suspicious devices is detected, the data flows of these devices are diverted to a quarantine network slice in which fine-grained detection, based on DPI, can be performed to set apart malicious devices from legitimate ones. By isolating suspicious devices, the operator can ensure that the QoS requirements of legitimate devices with a regular behaviour are fulfilled, whereas the performance of suspicious devices can still be guaranteed subject to the resource constraints in the quarantine slice. As in-depth traffic analysis is only performed when an abnormal behaviour is detected and only over a reduced subset of IoT traffic, the computational requirements of this detection solution is lower than previous solutions in which all the data flows are continuously sampled and deeply analyzed.

## 3. Proposed Solution

We propose an architecture based on Network Slicing for quarantining malicious IoT devices in mobile networks. The solution relies on NFV for creating multiple Network Slices with independent user-plane resources and on SDN for configuring the traffic flows in the network and redirecting each device seamlessly to the proper network slice. Furthermore, the proposed architecture is interoperable with 3GPP entities and therefore can be easily integrated in existing deployments.

Figure 1 shows a high-level overview of the proposed architecture. Each IoT device corresponds to a User Equipment (UE) in the mobile network. The UEs are connected to the Radio Access Network (RAN) through the base stations (i.e., eNodeB in 4G, gNodeB in 5G), which provide wireless connectivity to the IoT devices. In turn, the RAN is connected to the Core Network (CN) through an SDN network, consisting of at least one SDN switch in the user-plane that is managed by a logically centralized SDN controller. In the CN, our proposed architecture contains at least two network slices: an IoT Network Slice (IoT NS) for handling the traffic of legitimate IoT devices connected to the network and a Quarantine Network Slice (Quaratine NS), which is used for performing a deep inspection of potentially malicious IoT devices. Note that this approach can be generalized to more slices for IoT or other types of traffic. Moreover, the lifecycle of these network slices can be dynamically managed by leveraging NFV technology.

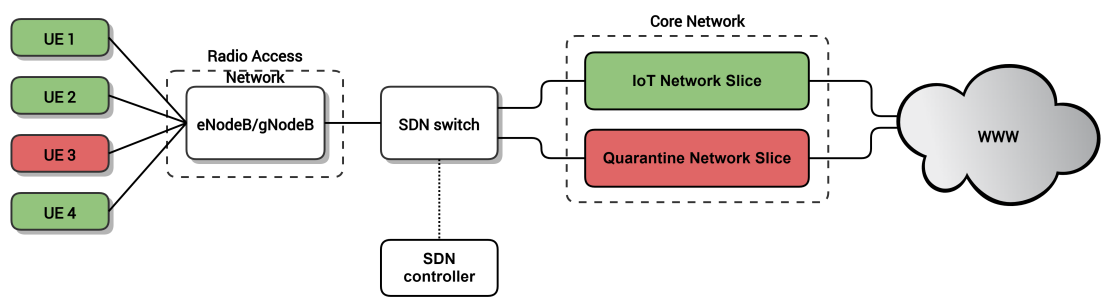

**Figure 1.** Proposed quarantine architecture.

Using this architecture, the SDN network interconnecting the RAN and the CN can assign UEs dynamically to the slices. This way, IoT devices are initially attached to the IoT NS, which provides full access to the network services the IoT devices require. An SDN application running on top of the SDN controller is in charge of identifying potentially malicious nodes and reassigning them to the Quarantine NS for further inspection. After a deep inspection of the traffic in the Quaratine NS, a device can be deemed malicious in which case its traffic is blocked or, conversely, it can be considered legitimate and consequently moved back to the IoT NS.

Overall, the proposed quarantine architecture employs a two-stage system for detecting malicious IoT devices: The first stage performs lightweight and coarse-grained early detection, which can involve some legitimate devices, while the second stage is a fine-grained in-depth traffic analysis for individually identifying malicious devices.

This architecture presents two major challenges that must be addressed: the detection of potentially malicious devices by the SDN controller and the dynamic reassignment of these devices to the different slices. On the one hand, the detection of potentially malicious nodes can be addressed by an SDN application monitoring the SDN flow rules installed on the SDN switches. As we show in this paper, the accuracy of this approach depends on the granularity of the flow rules in the SDN network. On the other hand, the dynamic reassignment of nodes to different slices can be achieved by redirecting their user-plane traffic to the desired slice using the SDN network. Moreover, this requires maintaining a state replication of the user-plane context among the different slices, so that the traffic of any device can be properly handled by any of them.

*3.1. Detection of Potential Malicious Devices*

An SDN application running on top of the SDN controller periodically samples the flow tables of the SDN switches, obtaining the numbers of packets and bytes transmitted in the sampling period for each flow rule in each SDN switch. Since the number of users connected to each base station can be up to 300,000 [28], installing a flow rule per each UE in each SDN switch is not feasible, because the tables of an SDN switch can efficiently support up to 4000 flow rules [29]. Therefore, in our solution each flow rule in the SDN switch corresponds to an aggregation of UEs. Then, the flows are the elements that the SDN application monitors and assigns to the slices. Consequently, the SDN application will detect suspicious flows, rather than malicious individual devices (note that we follow SDN terminology, in which the term flow denotes a sequence of packets that share the same set of packet header values).

A flow will be considered suspicious when its throughput exceeds a certain threshold. Thus, this approach requires knowing the typical traffic in a flow without malicious devices. Fortunately, there are several models for the behavior of IoT devices, and in particular their traffic patterns. IoT sensors are often sending data in a deterministic periodic manner [30–32], and other authors have also proposed deep learning methods for IoT traffic load prediction [33] for IoT devices with more complex communication patterns. Malicious devices usually perform scanning operations, attempts to infect other devices, and DDoS attacks (currently most IoT botnets are used for DDoS attacks) [34,35], thus increasing the regular traffic load.

An undesired side-effect of this aggregation is that legitimate devices may be temporarily assigned to the Quarantine NS due to the aggregation of nodes inside a flow. Nevertheless, those devices will not lose connectivity nor session continuity while they are diverted to the Quarantine NS, and they will be returned back to the IoT NS after actual malicious nodes are identified.

Another side-effect of the aggregation is that some flows containing malicious devices may not be properly detected if their activity does not exceed the threshold, so that they will never be sent to the Quarantine NS. However, although it is desirable to detect all suspicious devices, they will cause no harm if they are few. Furthermore, this problem can be alleviated by simply setting lower thresholds for tagging a flow as suspicious. These threshold values will determine the number of flows that are sent to the Quarantine NS.

*3.2. Dynamic Reassignment of Devices to Slices*

The dynamic reassignment of UEs to slices can be carried out in a transparent and seamless way both to the devices and to the mobile network elements by extending our previous work for transparent session and service continuity in dynamic Multi-Access Edge Computing (MEC) [36] to Network Slicing. This solution is valid both for 4G and 5G networks and consists in replicating the state of the user-plane functions to the different slices through an SDN application and then configuring the flow rules in the SDN switch to forward the traffic of each UE to the desired slice. The original solution described in [36] is intended for dynamic MEC, i.e., the traffic is originally forwarded to the core network and at some point, user-plane resources are dynamically deployed at the edge and the user-plane context is replicated from the core to the edge, before the traffic of the UE is diverted to the edge resources. In this paper, we generalize that solution to cover the case of Network Slicing, where both slices are deployed at the core network and the state replication solution is used for continuously maintaining state consistency between the two slices, and per-user traffic redirection is leveraged for transparently moving IoT devices from the IoT NS to the Quarantine NS and vice versa as needed.

## 4. Results

This section evaluates the proposed architecture for quarantining malicious IoT devices. We analyze the solutions presented for each of the challenges that we have identified. First we evaluate the detection of potentially malicious devices through a set of simulations and then analyze the performance of the dynamic reassignment of devices to slices in a real implementation.

*4.1. Detection*

We conducted simulations to evaluate the performance of the solution for detecting IoT DDoS attacks in mobile networks described in Section 3.1. The objective of these simulations was to estimate the probability of detecting malicious nodes within a set of devices corresponding to the same flow in the routing table of an SDN switch, which are perceived by the operator as a single unit, based on the comparison between issued and expected traffic of the devices in different scenarios.

To represent a realistic scenario and for simplicity the category of factory automation devices in [30] was considered in the simulations, as reproduced in Table 1. The behaviour of the corresponding IoT devices is highly predictable, since they send data frames with the same format and length at fixed periodicity [31,32]. This allows identifying malicious behavior by simply analyzing traffic deviations from the expected pattern. Our approach could be easily extended to other categories, even to categories identified in real time with Machine Learning (ML) or other techniques [37].

Table 1. Network characteristics of IoT devices in factory automation.

| Use Case | Latency (ms) | Reliability (PLR) | Transmission Frequency (ms) | Frame Data Size (bytes) | Device Density (Devices/m$^3$) | Communication Range (m) | Mobility (km/h) |
|---|---|---|---|---|---|---|---|
| Manufacturing cell | 5 | $10^9$ | 50 | 15 | 0.33 to 3 | 50 to 100 | <30 |
| Machine tools | 0.25 | $10^9$ | 0.5 | 50 | 0.33 to 3 | 50 to 100 | <30 |
| Printing machines | 1 | $10^9$ | 2 | 30 | 0.33 to 3 | 50 to 100 | <30 |
| Packaging machines | 25 | $10^9$ | 5 | 15 | 0.33 to 3 | 50 to 100 | <30 |

Different scenarios were simulated, each of them characterized by the following input parameters:

- $n$: Number of devices per flow.
- $p_m$: Probability of a device to be malicious (independent of other devices).
- $f_m$: Ratio between the transmission frequency of a malicious device and a legitimate device.
- $t_r$: Ratio between the threshold throughput, from which flows are detected as suspicious, and the volume of throughput expected for the current scenario.
- $p_t(i)$: Probability of a device to be of type $i$ (independent of other devices).
- $r(i)$: Data rate of type $i$ legitimate devices.
- $s_p$: Sampling period of the SDN application.
- $t_p(i)$: Transmission period of a type $i$ legitimate device.

We performed $10^5$ simulation rounds for each scenario, resulting in average percentages of malicious along with legitimate devices marked as suspicious under the conditions of each case. A custom simulator was developed to analyze the aforementioned scenarios. The source code of this tool has been released as open-source and is publicly available at [38].

On each round $n$ factory automation devices generate an aggregated flow that matches an entry in the flow table in an access SDN switch. The behaviour of each node is defined by its device type, which is selected by randomly picking one of the categories in Table 1 ({0, 1, 2, 3}) following a discrete uniform distribution (i.e., $p_t(i) = 0.25 \ \forall i \in \{0, 1, 2, 3\}$). Then, malicious devices were independently tagged according to a Bernoulli Be($p_m$) distribution, and configured to increase $f_m$ times the transmission frequency of legitimate devices to emulate a DDoS attack [34,39].

To simulate the traffic, the devices in the aggregated flow transmitted independently and asynchronously in an ideal non-saturated channel where data frames were always transmitted successfully. To emulate this behaviour, the arrival time of the first transmission since the beginning of the simulation followed a uniform U(0,$t_p(i)$) distribution. Under these assumptions, the aggregated data rate was measured as the sum of the data traffic generated by the aggregated flow, by counting the number and length of frames transmitted by each device (which depends on their transmission frequency and frame data size) during the SDN application sampling period $s_p$. In a real setup, this procedure would be performed by an SDN application by directly sampling the flow tables of the SDN switches every $s_p$ seconds to obtain the throughput. The expected aggregated traffic was computed as the sum of the average data rates of the devices. If the aggregated data rate exceeded the expected one in a ratio higher than $t_r$, the flow was tagged as suspicious and their devices diverted to the Quarantine NS for further inspection.

Figure 2 illustrates the percentage of devices that were quarantined (that is, the number of devices in flows with abnormal traffic) as a function of the probability of a device to be compromised or malicious. Traffic was considered anomalous if it exceeded the expected regular rate by 1% (i.e., $t_r = 1.01$). Malicious devices transmitted 100 times more data than legitimate ones (i.e., $f_m = 100$) and the SDN application sampling period $s_p$ was set to 1 s. In this scenario, recall is significant even for small populations of malicious devices in large aggregated flows. In fact, more than 75% of the malicious devices are quarantined for a $p_m$ higher than 0.001 for any number of devices per aggregated flow. As expected, recall increases with $p_m$. All the devices in a suspicious aggregated flow, which may include many legitimate devices, are quarantined. Thus, the number of legitimate nodes diverted to the Quarantine NS increases with the number of aggregated devices ($n$). The fewer devices are aggregated per flow, the greater the accuracy of the system.

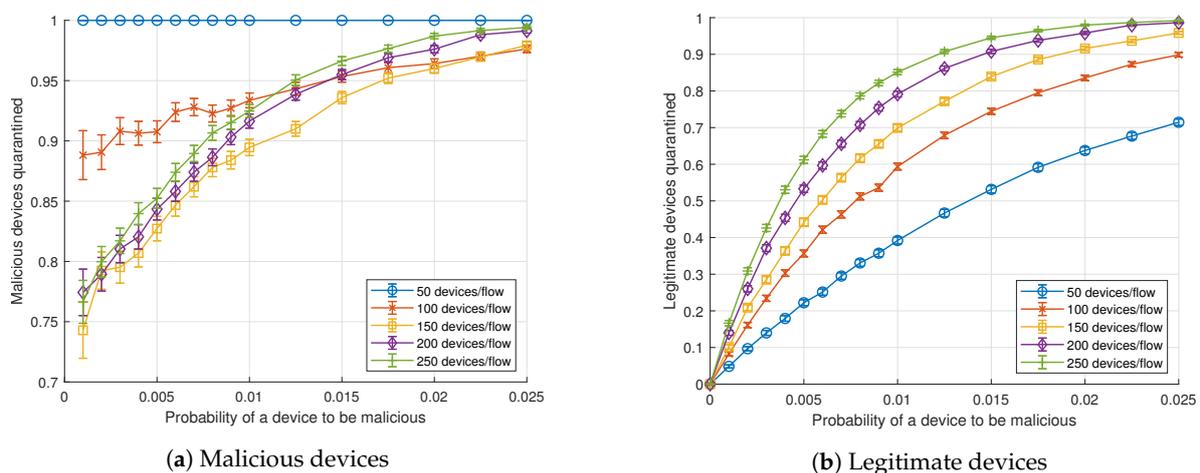

(**a**) Malicious devices　　　　(**b**) Legitimate devices

**Figure 2.** Ratio of devices quarantined as a function of the probability of a device to be malicious, $p_m$ ($s_p = 1\,\text{s}$, $t_r = 1.01$, $f_m = 100$), for $n \in \{50, 100, 150, 200, 250\}$.

A closer observation of Figure 2a reveals that the effect of the number of devices aggregated in a flow ($n$) on the detection of malicious devices in the simulated scenario is not intuitive. Any malicious device is detected for $n = 50$ devices/flow, but then the ratio of malicious devices detected decreases as the number of devices per flow grows up to 150. After that value, detection improves as $n$ increases. The cause of this effect is as follows. For a low value of $n$, the probability that there is no more than one malicious device within the aggregated flow is high (note that Figure 2a only evaluates flows containing malicious devices). However, within such reduced set of nodes, the malicious device is easily detected. At a certain $n$, flows containing only a manufacturing cell type malicious node (the category with the lowest rate) begin to go unnoticed and, consequently, detection drops. But, on the other side, increasing the size of the aggregated flow also decreases the probability of flows containing only malicious nodes of manufacturing cell type, thus raising the detection (other types of malicious nodes are easier to detect because of their higher rate).

As an example to better understand this effect, we have analytically studied the probability of detecting a flow which contains malicious devices as a function of $n$ in the scenario described in Figure 2a for a $p_m = 0.01$. Figure 3 shows the probability of detecting a flow which contains malicious devices in this scenario, and the contribution to this probability by flows which contain some malicious device of a type different than manufacturing cell. The first we can observe is that the values at 50, 100, 150, 200 and 250 devices per flow match with the experimental results in Figure 2a for a $p_m = 0.01$. As we can see, the probability of detecting a flow with malicious devices is 1 for $n < 70$ devices per flow. Then, this value starts decreasing because the probability of detecting a flow containing only malicious devices of manufacturing cell type decreases. On the other hand, the contribution of flows with malicious devices raises with $n$, validating our previous arguments. Analytical formulas used for studying this scenario are elaborated in Appendix A.1.

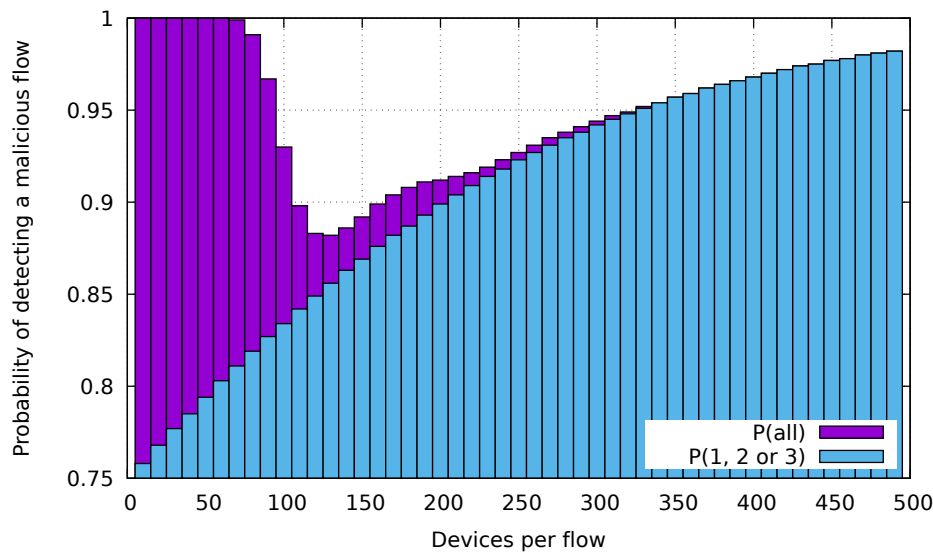

**Figure 3.** Probability of detecting a flow which contains malicious devices (given by Equation (A1)) and the contribution of flows which contain malicious devices of type different than manufacturing cell (given by Equation (A2)) for $s_p = 1$ s, $t_r = 1.01$, $f_m = 100$ and $p_m = 0.01$.

Figure 4 shows the effect of the threshold ratio ($t_r$) on the detection of malicious devices, for $n = 100$ devices per aggregated flow, $s_p = 1$ s, $f_m = 100$ and different $p_m$. The higher the threshold, the more malicious devices get unnoticed. For low values of $n$ and $p_m$, there is a low probability for two or more malicious nodes of the same device to exist within the same aggregated flow. Under these conditions, detection drops around specific threshold ratio values for a device type of $x$, given by Equation (1).

$$threshold\_detection\_limit(x) = 1 + \frac{r(x) \times (f_m - 1)}{\sum_{i=0}^{3} r(i) \times p_t(i) \times n} \quad (1)$$

That is, there is a value for the threshold from where detection decays. Such threshold is high enough to hide the traffic of a specific type.

In the same way, the number of legitimate users quarantined also decreases with the threshold, but at a slower pace, as depicted in Figure 4b. In fact, as the threshold increases, the number of legitimate nodes quarantined is $1 - (1 - p_m)^n$ times lower than the malicious nodes quarantined (being $1 - (1 - p_m)^n$ the probability that there are malicious nodes in the aggregated flow).

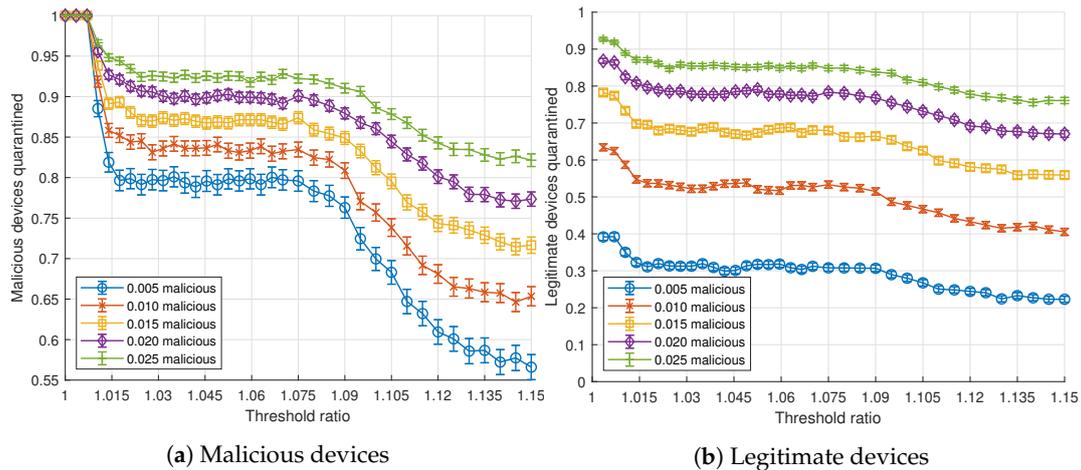

(**a**) Malicious devices  (**b**) Legitimate devices

**Figure 4.** Ratio of devices quarantined as a function of the detection threshold ratio, $t_r$ ($n = 100$, $s_p = 1$ s, $f_m = 100$) for $p_m \in \{0.005, 0.010, 0.015, 0.020, 0.025\}$.

Figure 5 shows the effect of the ratio between the transmission frequencies of malicious and legitimate devices ($f_m$) on detection, for $n = 100$ devices per aggregated flow, $s_p = 1$ s, a $t_r = 1.01$ and different $p_m$. Even for low deviations of the $f_m$, and for a low $p_m$, the detection recall for malicious nodes is high. For example, for a transmission frequency increase of $\times 20$, over 75 % of malicious devices are properly quarantined, whereas a transmission frequency above $\times 125$ yields a 100 % proper quarantining of all malicious devices in all scenarios. However, as already stated, since all the devices within an aggregated flow are quarantined, higher recalls involve more legitimate devices being sent to the Quarantine NS. Nevertheless, the amount of quarantined legitimate nodes also converges at $f_m > 125$. This is explained by the fact that, with this configuration, all flows containing at least one malicious device are correctly quarantined, and no flow just composed of legitimate nodes is quarantined. Therefore, the total number of quarantined flows will be bounded by the amount of malicious devices in the network, since in the worst-case scenario each malicious device belongs to a different aggregated flow in the SDN switch. However, even assuming a uniform distribution of devices among flows as in our model, the average percentage of saturation will be lower. Specifically, the expected number of malicious flows is given by the probability that a flow contains malicious devices: $1 - (1 - p_m)^n$. Table 2 shows the expected numbers of malicious flows. As we can observe, these values match the results in Figure 5b for $f_m > 125$, as expected.

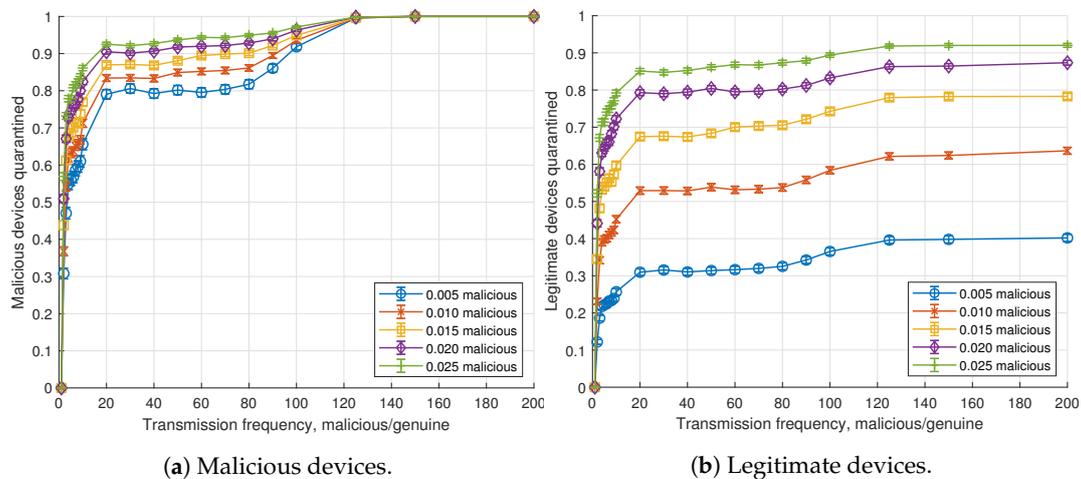

(**a**) Malicious devices.  (**b**) Legitimate devices.

**Figure 5.** Ratio of devices quarantined as a function of the transmission frequency of malicious devices, $f_m$ ($n = 100$, $s_p = 1$ s, $t_r = 1.01$) for $p_m \in \{0.005, 0.010, 0.015, 0.020, 0.025\}$.

Table 2. Expected number of malicious flows vs. malicious devices.

| Malicious Devices (%) | Expected Number of Malicious Flows (%) |
|---|---|
| 0.5 | 39.4 |
| 1.0 | 63.4 |
| 1.5 | 77.9 |
| 2.0 | 86.7 |
| 2.5 | 92.0 |

The effect of the SDN application sampling period ($s_p$) was also analyzed through simulations with values ranging from 200 ms (as the maximum transmission period of the devices was 50 ms) to 1.5 s. Results showed no effect on the detection of flows containing malicious devices nor on the quarantine of legitimate nodes. This is true provided that the sampling period of the SDN application is higher than the transmission period of the devices, detection capabilities are not modified. Indeed values lower than 1 s for the sampling period will not be valid in practice due to the overhead introduced in the SDN controller and sampling the flow tables of the switches. On the other hand, the value of the sampling period measures the response time needed for identifying suspicious flows. Thus a value of $s_p = 1$ s establishes an acceptable trade-off.

The simulation outputs were matched against the analytical models provided in Appendix A.2.

### 4.2. Dynamic Reassignment

In this section we evaluate the proposed mechanism for dynamic reassignment of IoT devices to slices, which, as previously said, is an extension of our solution for dynamic MEC in [36]. We analyzed the performance of the proposed solution through an experiment in a real implementation.

Due to the lack of open-source 5G network implementations, we decided to conduct our experiments in a 4G network. We built a testbed based on the architecture in Figure 6 using the 4G stack provided by OpenAirInterface [40] software, both for the RAN and for the Core Network; Open vSwitch [41] for the SDN switch interconnecting the RAN and the Core Network; and the Open Network Operating System (ONOS) [42] for the SDN controller.

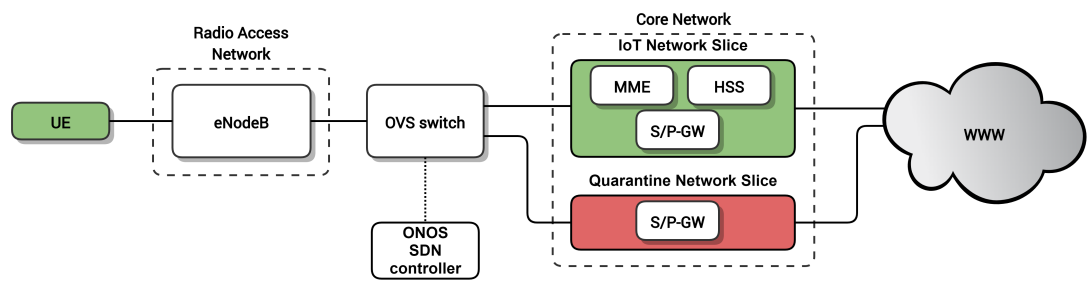

**Figure 6.** Architecture of the setup used for the experiments.

In detail, the elements used in the testbed were:

- UE: Huawei E3372h-153 Long Term Evolution (LTE) modem connected to a laptop (Intel®Core™i7 4510U CPU @ 2.0 GHz).
- Radio Access Network: eNodeB software implementation provided by OpenAirInterface running on a commodity PC (Intel®Core™i7-6700 CPU @ 3.40GHz) connected through USB-3 to an Ettus USRP B210 SDR board.
- Core Network: The network slices of the Core Network were deployed as Docker containers on top of an Intel®Xeon®CPU E5-2603 v3 @ 1.60GHz.

- IoT NS: Three Docker containers implementing the main Evolved Packet Core (EPC) entities using OpenAirInterface software, Home Subscriber Server (HSS), Mobility Management Entity (MME) and Serving/Packet Data Network (PDN) Gateway (S/P-GW).
- Quarantine NS: One Docker container implementing an OpenAirInterface S/P-GW.

- SDN switch: Open vSwitch bridge with extended General Packet Radio Service (GPRS) Tunneling Protocol (GTP) encapsulating and decapsulating capabilities ( [43,44]) created on top of an Accelerated Processing Unit (APU) from PC engines(https://www.pcengines.ch/apu.htm).
- SDN controller: ONOS SDN controller deployed as a Docker container in an Intel®Xeon®CPU E5-2603 v3 @ 1.60GHz. The SDN applications for moving the devices/UEs were implemented on top of ONOS as described in [36]. These SDN applications allow for replicating the state of a S/P-GW into another one and diverting the traffic of specific users to the desired S/P-GW.

The experiment we conducted on this testbed was to measure the time required for reassigning a device/UE from the IoT NS to the Quarantine NS ($\Delta$). This time includes:

1. Dynamic creation of the Quarantine NS: Deployment of the S/P-GW Docker container in the Core Network ($\Delta_1$).
2. S/P-GW application initialization ($\Delta_2$).
3. State replication of the UE context in the S/P-GW of the IoT NS to the S/P-GW in the Quarantine NS ($\Delta_3$).
4. Data-plane reconfiguration of flow rules in the SDN switch for forwarding the traffic of the device/UE to the Quarantine NS ($\Delta_4$).

The first step was performed by the Docker engine at the Core Network whereas the last two steps were performed by an SDN application running on top of the ONOS SDN controller. Note that these steps must proceed sequentially for a successful reassignment without service interruption. Thus, $\Delta = \Delta_1 + \Delta_2 + \Delta_3 + \Delta_4$.

The methodology we followed for the experiment consisted in first deploying the SDN controller and the IoT NS in the Core Network and then connecting the eNodeB for a functional 4G network. Then, one static UE is connected to the 4G network generated using the OpenAirInterface software. The UE is intially attached to the IoT NS. Then, we complete different steps and measure their times: Instruct the Docker engine to deploy a S/P-GW container and wait for the container to be deployed, wait for the S/P-GW container to be properly initialized, instruct the SDN application running on top of the ONOS SDN controller to replicate the context of the UE in the S/P-GW of the IoT NS to the new S/P-GW corresponding to the Quarantine NS and finally instruct the SDN controller to reconfigure the flow rules of the SDN switch to send the traffic of the UE to the S/P-GW in the Quarantine NS. We verified that the UE continued attached to the network during the whole process.

For complete clean-up of the experiment before subsequent executions, we disconnected the UE, stopped the eNodeB and removed all Docker containers corresponding both to the Core Network entities of both slices and also the SDN controller.

This experiment was run 100 times for statistical significance.

Table 3 shows the time required for reassigning a device from the IoT NS to the Quarantine NS and the individual contributions to this time, averaging 100 independent executions. The first fact we can notice is the stability of the results through the different executions. Their variability is very low except for $\Delta_4$. We can see that the total time required for moving one device to the Quarantine NS ($\Delta$) is about 1 s. An inspection of this time reveals that the major contribution came from the deployment of the S/P-GW Docker container ($\Delta_1$) and the initialization of the S/P-GW software ($\Delta_2$), which took about 600 ms and 400 ms, respectively. On the other hand, S/P-GW state replication ($\Delta_3$) was an order of magnitude lower requiring less than 40 ms, while data-plane reconfiguration ($\Delta_4$) took less than 10 ms. Note that the results for $\Delta_4$ are in line with recent measurements of flow setup times in SDN-enabled switches [45].

Table 3. Device/UE reassignment time and its components (averaging 100 independent executions).

|  | Mean | Median | Minimum | Maximum | 95-th Percentile |
|---|---|---|---|---|---|
| $\Delta$ (ms) | 989.69 | 992.20 | 918.32 | 1057.86 | 1027.05 |
| $\Delta_1$ (ms) | 570.11 | 571.48 | 505.27 | 630.98 | 606.04 |
| $\Delta_2$ (ms) | 381.68 | 381.46 | 361.89 | 397.13 | 392.32 |
| $\Delta_3$ (ms) | 33.92 | 33.71 | 32.91 | 37.67 | 35.38 |
| $\Delta_4$ (ms) | 3.98 | 3.84 | 0.61 | 9.98 | 6.03 |

This reassignment time in the order of one second determines the reaction speed of the first stage of our proposed solution. Since the legitimate devices redirected to the Quarantine NS will be reassigned back to the IoT NS, the time this takes is also relevant for our solution. However, note that this action simply requires a data plane reconfiguration, because the IoT NS is already deployed, the S/P-GW is initialized and it has the required state. As a result, moving a device back to the IoT NS is a lightweight process taking less than 10 ms as measured in $\Delta_4$ in Table 3.

Furthermore, the process of moving a device to the Quarantine NS as described is fully reactive, comprising the deployment of new data plane resources on demand. As analyzed, this part is indeed the most time consuming in the whole reassignment process. We thus propose as an alternative a proactive approach for reducing reassignment time: Instead of creating the S/P-GW in the Quarantine NS as part of the reassignment process, a S/P-GW can be proactively created and initialized simultaneously to the IoT NS. This way, the reassignment time ($\Delta$) will correspond to the contributions of replicating the state of the device in the IoT S/P-GW ($\Delta_3$) and reconfiguring the data plane ($\Delta_4$), which adds up to less than 50 ms on the worst case. This time can be further reduced by proactively replicating the state of the devices in the IoT S/P-GW to the Quarantine S/P-GW, maintaining state consistency between both entities. In this latest configuration, $\Delta$ is fully minimized for the proposed architecture, yielding a value of less than 10 ms for $\Delta_4$. It is important to bear in mind the trade-off introduced by the proactive approach in terms of increased resource usage due to the permanent deployment of the S/P-GW in the Quarantine NS and the control information overhead by maintaining full state consistency between the S/P-GW entities.

## 5. Conclusions

Massive IoT networks are challenging for 5G networks, not only in terms of scalability, but also of security. However, conventional security solutions have focused on eMBB scenarios and do not fit well into massive IoT communications, due to their huge density of devices. Hence, operators need tailored solutions for this use case.

In this paper we propose an architecture based on Network Slicing, NFV and SDN for detecting malicious devices and mitigating their impact in massive IoT scenarios, which can be easily integrated into any existing deployment due to its interoperability. The solution in this architecture is a two-step process: the first step, which we analyze in depth in this paper, performs a coarse-grained lightweight detection of potentially malicious devices through SDN techniques. In the second step, suspicious devices can then be seamlessly redirected to a Quarantine Network Slice for fine-grained in-depth inspection.

We have validated the flow analysis mechanism proposed through simulations and evaluated the performance of the redirection with a real testbed implementation. The results of the simulations reveal a trade-off between the probability of successful identification of malicious devices and the number of legitimate devices that are sent to the quarantine slice, which depends on the number of devices in the aggregated flows of the SDN switches and the detection threshold. These parameters can be configured in the SDN application to fine-tune the desired performance of the application. On the other hand, the implementation of the proposed architecture in a real testbed allowed us validating its feasibility and also measuring that moving a device to the quarantine slice takes about 1 s when the quarantine slice is reactively deployed on demand. The results also show that this time can be further reduced to less than 10 ms if the quarantine slice is proactively deployed.

Our proposed architecture is particularly efficient in scenarios with a large number of IoT devices because it does not need to inspect all the traffic in depth, but just the devices corresponding to aggregated SDN flows with suspicious behavior. Moreover, unlike conventional mechanisms, our solution does only trigger an in-depth inspection of the traffic in the event of a potentially malicious attack, thus reducing the computational load in the operator network. This architecture is specially useful and adequate for environments with a small number of malicious devices. Our experiments and models demonstrated how malicious devices were quarantined for further analysis even for low ratios of compromised devices or devices transmitting small amounts of information.

Although our method also quarantines legitimate devices that are in the same flow of malicious devices, their communications are kept active while they are being inspected and also when legitimate devices are returned to the original slice, making our approach transparent and non intrusive.

As future work we plan to implement and evaluate a mechanism for performing in-depth traffic inspections in the quarantine slice.

**Author Contributions:** Conceptualization, F.G.-C. and F.J.G.-C.; methodology, F.J.G.-C.; software, D.C.-V. and P.F.-F.; validation, D.C.-V., P.F.-F. and F.G.-C.; formal analysis, P.F.-F.; investigation, D.C.-V. and P.F.-F.; resources, F.G.-C.; data curation, D.C.-V.; writing—original draft preparation, D.C.-V. and P.F.-F.; writing—review and editing, F.G.-C. and F.J.G.-C.; visualization, D.C.-V.; supervision, F.G.-C. and F.J.G.-C.; project administration, F.G.-C. and F.J.G.-C.; funding acquisition, F.G.-C., F.J.G.-C. and P.F.-F.. All authors have read and agreed to the published version of the manuscript.

**Funding:** This research was partially funded by a "la Caixa" Foundation (ID 100010434) fellowship (LCF/BQ/ES18/11670020); Ministerio de Economía, Industria y Competitividad, Spain (TEC2016-76465-C2-2-R); and Xunta de Galicia (GRC2018/053,ED341D-R2016/012).

**Conflicts of Interest:** The authors declare no conflict of interest. The funders had no role in the design of the study; in the collection, analyses, or interpretation of data; in the writing of the manuscript, or in the decision to publish the results.

## Abbreviations

The following abbreviations are used in this manuscript:

| | |
|---|---|
| 3GPP | 3rd Generation Partnership Project |
| APU | Accelerated Processing Unit |
| CN | Core Network |
| DDoS | Distributed Denial of Service |
| DPI | Deep Packet Inspection |
| eMBB | enhanced Mobile Broadband |
| EPC | Evolved Packet Core |
| GPRS | General Packet Radio Service |
| GTP | GPRS Tunneling Protocol |
| HSS | Home Subscriber Server |
| IoT | Internet of Things |
| LTE | Long Term Evolution |
| MEC | Multi-Access Edge Computing |
| MME | Mobility Management Entity |
| NFV | Network Function Virtualization |
| NS | Network Slice |
| ONOS | Open Network Operating System |
| PDN | Packet Data Network |
| QoS | Quality of Service |
| RAN | Radio Access Network |
| SDN | Software-Defined Networking |
| SLA | Service Level Agreement |
| S/P-GW | Serving/PDN Gateway |
| UE | User Equipment |

**Appendix A. Analytical Model**

*Appendix A.1. Probability of a Flow with Malicious Devices to Be Detected in Specific Scenarios*

This appendix develops a formula tailored to scenarios where all flows containing some malicious device of type different than 0 are detected. In particular, these formulas have been used to analyze the scenario with the following configuration arised from Figure 2a:

- $p_m = 0.01$
- $f_m = 100$
- $t_r = 1.01$
- $p_t(i) = 0.25 \; \forall i$
- $r(0) = 2.4 \, \text{kbit/s}, r(1) = 800 \, \text{kbit/s}, r(2) = 120 \, \text{kbit/s}$ and $r(3) = 24 \, \text{kbit/s}$.
- $n < 500$

The model builds on the fact that for this configuration, a flow containing some malicious device of type different than manufacturing cell will always be detected.

Let us define the following list of events to be used in the formulas:

- *A*: A flow is detected as malicious.
- *B*: A flow contains malicious devices.
- *C*: A flow contains some malicious device of type different than 0 (manufacturing cell).
- *D(i)*: A flow contains exactly *i* malicious devices of type 0 and no other malicious device.

$$P(A \mid B) = P((A \cap C) \mid B) + P((A \cap C') \mid B) \tag{A1}$$

where $C'$ denotes the complementary event of C and equality holds due to the law of total probability.

$$P((A \cap C) \mid B) = \frac{P(A \cap C \cap B) \times P(B \mid (A \cap C))}{P(B)} = \frac{P(A \cap C \cap B)}{P(B)} \tag{A2}$$

where the first step holds due to Bayes' theorem and the second step holds because $P(B \mid (A \cap C)) = 1$ since flows detected as malicious always contain malicious devices (i.e., flows without malicious devices are never detected as malicious).

$$P((A \cap C') \mid B) = \frac{P(A \cap C' \cap B) \times P(B \mid (A \cap C'))}{P(B)} = \frac{P(A \cap C' \cap B)}{P(B)} \tag{A3}$$

where the first step holds due to Bayes' theorem and the second step holds because $P(B \mid (A \cap C')) = 1$ since flows detected as malicious always contain malicious devices (i.e., flows without malicious devices are never detected as malicious).

$$P(B) = P\left(\text{B}(n, p_m) > 0\right) = 1 - (1 - p_m)^n \tag{A4}$$

$$P(A \cap C \cap B) = P(A \cap C) = P(A \mid C) \times P(C) = P(C) \tag{A5}$$

where the first step holds because $C \subset B$ and thus $C \cap B = C$ (i.e., a flow which contains some malicious device of type different than 0 always contains malicious devices).

$$P(C) = P\left(\text{B}\left(n, p_m \times \sum_{k=0}^{3} p_t(k)\right) > 0\right) = 1 - \left(1 - \left(p_m \times \sum_{k=0}^{3} p_t(k)\right)\right)^n \tag{A6}$$

$$P(A \mid C) = 1 \tag{A7}$$

where this holds from the fact that a throughput excess of $\min_{i\neq 0}(r(i))$ always exceeds the threshold ratio $t_r$ for our input parameters of $p_m = 0.01$, $f_m = 100$, $t_r = 1.01$ $n < 500$, $p_t(i) = 0.25$ $\forall i$, $r(0) = 2.4\,\text{kbit/s}$, $r(1) = 800\,\text{kbit/s}$, $r(2) = 120\,\text{kbit/s}$ and $r(3) = 24\,\text{kbit/s}$.

$$P(A \cap C' \cap B) = \sum_{i=0}^{n} P(A \cap C' \cap B \cap D(i)) = \sum_{i=1}^{n} P(A \cap D(i)) = \sum_{i=1}^{n} P(A \mid D(i)) \times P(D(i)) \quad \text{(A8)}$$

where the first step holds for the law of total probability and the second one holds because $C' \cap B \cap D(i) = D(i)$ $\forall i \neq 0$ and $0$ for $i = 0$.

$$P(D(i)) = \binom{n}{i}(p_m \times p_t(0))^i \times (1 - p_m)^{n-i} \quad \text{(A9)}$$

$$\begin{aligned}
P(A \mid D(i)) = & \sum_{n_0=0}^{n} \sum_{n_1=0}^{n-n_0} \sum_{n_2=0}^{n-n_0-n_1} \binom{n}{n_0}\binom{n-n_0}{n_1}\binom{n-n_0-n_1}{n_2} \\
& \times (p_t(0))^{n_0} \times (p_t(1))^{n_1} \times (p_t(2))^{n_2} \times (p_t(3))^{n-n_0-n_1-n_2} \\
& \times \left[\frac{(n_0 \times r(0) + n_1 \times r(1) + n_2 \times r(2) + (n-n_0-n_1-n_2) \times r(3)) \times (t_r - 1)}{(f_m - 1) \times r(0)} < i\right]
\end{aligned} \quad \text{(A10)}$$

where [expression] notation refers to Iverson brackets (i.e., [expression] is 1 if expression is true and 0 otherwise).

*Appendix A.2. Probability of a Flow with Malicious Devices to Be Correctly Detected (General Formula)*

This appendix develops a general formula for the probability of flow being detected as malicious provided that it contains malicious devices. It has been used for validating the results provided by the simulations.

Let us define the following list of events to be used in the formulas:

- $A$: A flow is detected as malicious.
- $B$: A flow contains malicious devices.
- $E(x)$: A flow generates $x$ excess throughput due to malicious devices.

$$P(A \mid B) = \sum_{\forall x=(t_r-1)\times \sum_{i=0}^{3}(n_i \times r(i)) \mid n_i \in \{0,\ldots,n\},\, \sum_{i=0}^{3} n_i \leq n} P((A \cap E(x)) \mid B) \quad \text{(A11)}$$

where this holds for the law of total probability.

$$\begin{aligned}
P((A \cap E(x)) \mid B) &= \frac{P(A \cap E(x)) \times P(B \mid (A \cap E(x)))}{P(B)} = \frac{P(A \cap E(x))}{P(B)} \\
&= \frac{P(A \mid E(x)) \times P(E(x))}{P(B)}
\end{aligned} \quad \text{(A12)}$$

where the first step holds due to Bayes' theorem, the second step holds because $P(B \mid (A \cap E(x))) = 1$ since flows detected as malicious always contain malicious devices (i.e., flows without malicious devices are never detected as malicious).

$$P(B) = P(\text{B}(n, p_m) > 0) = 1 - (1 - p_m)^n \quad \text{(A13)}$$

$$P(E(x)) = \sum_{n_0=0}^{n} \sum_{n_1=0}^{n-n_0} \sum_{n_2=0}^{n-n_0-n_1} \binom{n}{n_0} \binom{n-n_0}{n_1} \binom{n-n_0-n_1}{n_2}$$
$$\times (p_t(0))^{n_0} \times (p_t(1))^{n_1} \times (p_t(2))^{n_2} \times (p_t(3))^{n-n_0-n_1-n_2}$$
$$\times [(n_0 \times r(0) + n_1 \times r(1) + n_2 \times r(2) + (n - n_0 - n_1 - n_2) \times r(3)) \times (t_r - 1) = x] \quad (A14)$$

$$P(A \mid E(x)) = \sum_{n_0=0}^{n} \sum_{n_1=0}^{n-n_0} \sum_{n_2=0}^{n-n_0-n_1} \sum_{n_3=0}^{n-n_0-n_1-n_2} \binom{n}{n_0} \binom{n-n_0}{n_1} \binom{n-n_0-n_1}{n_2}$$
$$\times \binom{n-n_0-n_1-n_2}{n_3} \times p_m^{n_0+n_1+n_2+n_3} \times (1-p_m)^{n-n_0-n_1-n_2-n_3}$$
$$\times (p_t(0))^{n_0} \times (p_t(1))^{n_1} \times (p_t(2))^{n_2} \times (p_t(3))^{n_3}$$
$$\times [(n_0 \times r(0) + n_1 \times r(1) + n_2 \times r(2) + n_3 \times r(3)) \times (f_m - 1) > x] \quad (A15)$$

where [expression] notation refers to Iverson brackets.